\pdfoutput=1

\documentclass[12pt,a4paper]{article}

\usepackage{ifthen} 
\newboolean{pdflatex}
\setboolean{pdflatex}{true} 

\newboolean{articletitles}
\setboolean{articletitles}{true} 

\newboolean{uprightparticles}
\setboolean{uprightparticles}{false} 

\newboolean{inbibliography}
\setboolean{inbibliography}{false} 

\usepackage{microtype}
\usepackage{lineno}  
\usepackage{xspace} 

\usepackage{graphicx}  
\usepackage{color}
\usepackage{colortbl}
\graphicspath{{./figs/}} 

\usepackage{amsmath} 
\usepackage{amssymb}
\usepackage{amsfonts}
\usepackage{upgreek} 

\newcommand*\patchAmsMathEnvironmentForLineno[1]{%
\expandafter\let\csname old#1\expandafter\endcsname\csname #1\endcsname
\expandafter\let\csname oldend#1\expandafter\endcsname\csname
end#1\endcsname
 \renewenvironment{#1}%
   {\linenomath\csname old#1\endcsname}%
   {\csname oldend#1\endcsname\endlinenomath}%
}
\newcommand*\patchBothAmsMathEnvironmentsForLineno[1]{%
  \patchAmsMathEnvironmentForLineno{#1}%
  \patchAmsMathEnvironmentForLineno{#1*}%
}
\AtBeginDocument{%
\patchBothAmsMathEnvironmentsForLineno{equation}%
\patchBothAmsMathEnvironmentsForLineno{align}%
\patchBothAmsMathEnvironmentsForLineno{flalign}%
\patchBothAmsMathEnvironmentsForLineno{alignat}%
\patchBothAmsMathEnvironmentsForLineno{gather}%
\patchBothAmsMathEnvironmentsForLineno{multline}%
}

\usepackage{hyperref}    
\usepackage[all]{hypcap} 

\def\lhcb {\mbox{LHCb}\xspace}
\def\ux85 {\mbox{UX85}\xspace}

\def\lhc    {\mbox{LHC}\xspace}

\ifthenelse{\boolean{uprightparticles}}%
{
 
 \def\Pgamma      {\ensuremath{\upgamma}\xspace}

 \def\Pmu         {\ensuremath{\upmu}\xspace}

 \def\Ppi         {\ensuremath{\uppi}\xspace}

 \def\Pchi        {\ensuremath{\upchi}\xspace}                 
 \def\Ppsi        {\ensuremath{\uppsi}\xspace}

 \def\PDelta      {\ensuremath{\Delta}\xspace}                 
 \def\PXi      {\ensuremath{\Xi}\xspace}                 
 \def\PLambda      {\ensuremath{\Lambda}\xspace}                 
 \def\PSigma      {\ensuremath{\Sigma}\xspace}                 
 \def\POmega      {\ensuremath{\Omega}\xspace}                 
 \def\PUpsilon      {\ensuremath{\Upsilon}\xspace}                 
 
 \def\PB      {\ensuremath{\mathrm{B}}\xspace}                 
                  
 \def\PD      {\ensuremath{\mathrm{D}}\xspace}

 \def\PJ      {\ensuremath{\mathrm{J}}\xspace}                 
 \def\PK      {\ensuremath{\mathrm{K}}\xspace}

 \def\Pb      {\ensuremath{\mathrm{b}}\xspace}                 
 \def\Pc      {\ensuremath{\mathrm{c}}\xspace}                 
                  
 \def\Pe      {\ensuremath{\mathrm{e}}\xspace}

 \def\Pi      {\ensuremath{\mathrm{i}}\xspace}

 \def\Pp      {\ensuremath{\mathrm{p}}\xspace}

}
{
 
 \def\Pgamma      {\ensuremath{\gamma}\xspace}

 \def\Pmu         {\ensuremath{\mu}\xspace}

 \def\Ppi         {\ensuremath{\pi}\xspace}

 \def\Pchi        {\ensuremath{\chi}\xspace}                 
 \def\Ppsi        {\ensuremath{\psi}\xspace}                 
                  
 \mathchardef\PDelta="7101
 \mathchardef\PXi="7104
 \mathchardef\PLambda="7103
 \mathchardef\PSigma="7106
 \mathchardef\POmega="710A
 \mathchardef\PUpsilon="7107
                  
 \def\PB      {\ensuremath{B}\xspace}                 
                  
 \def\PD      {\ensuremath{D}\xspace}

 \def\PJ      {\ensuremath{J}\xspace}                 
 \def\PK      {\ensuremath{K}\xspace}

 \def\Pb      {\ensuremath{b}\xspace}                 
 \def\Pc      {\ensuremath{c}\xspace}                 
                  
 \def\Pe      {\ensuremath{e}\xspace}

 \def\Pi      {\ensuremath{i}\xspace}

 \def\Pp      {\ensuremath{p}\xspace}

}

\def\en         {\ensuremath{\Pe^-}\xspace}   
\def\ep         {\ensuremath{\Pe^+}\xspace}
 
\def\epem       {\ensuremath{\Pe^+\Pe^-}\xspace}

\def\mumu       {\ensuremath{\Pmu^+\Pmu^-}\xspace}

\def\g      {\ensuremath{\Pgamma}\xspace}

\def\cquark    {\ensuremath{\Pc}\xspace}
\def\cquarkbar {\ensuremath{\overline \cquark}\xspace}
\def\ccbar     {\ensuremath{\cquark\cquarkbar}\xspace}
\def\bquark    {\ensuremath{\Pb}\xspace}

\def\pion  {\ensuremath{\Ppi}\xspace}
\def\piz   {\ensuremath{\pion^0}\xspace}

\def\kaon  {\ensuremath{\PK}\xspace}
  \def\Kbar  {\kern 0.2em\overline{\kern -0.2em \PK}{}\xspace}

\def\Kz    {\ensuremath{\kaon^0}\xspace}
\def\Kzb   {\ensuremath{\Kbar^0}\xspace}
\def\KzKzb {\ensuremath{\Kz \kern -0.16em \Kzb}\xspace}
\def\Kp    {\ensuremath{\kaon^+}\xspace}
\def\Km    {\ensuremath{\kaon^-}\xspace}

\def\KpKm  {\ensuremath{\Kp \kern -0.16em \Km}\xspace}

  \def\Dbar    {\kern 0.2em\overline{\kern -0.2em \PD}{}\xspace}
\def\D       {\ensuremath{\PD}\xspace}

\def\Dz      {\ensuremath{\D^0}\xspace}
\def\Dzb     {\ensuremath{\Dbar^0}\xspace}
\def\DzDzb   {\ensuremath{\Dz {\kern -0.16em \Dzb}}\xspace}
\def\Dp      {\ensuremath{\D^+}\xspace}
\def\Dm      {\ensuremath{\D^-}\xspace}

\def\DpDm    {\ensuremath{\Dp {\kern -0.16em \Dm}}\xspace}

  \def\Bbar    {\kern 0.18em\overline{\kern -0.18em \PB}{}\xspace}

\def\jpsi     {\ensuremath{{\PJ\mskip -3mu/\mskip -2mu\Ppsi\mskip 2mu}}\xspace}
\def\psitwos  {\ensuremath{\Ppsi{(2S)}}\xspace}

\def\chiczero {\ensuremath{\Pchi_{\cquark 0}}\xspace}
\def\chicone  {\ensuremath{\Pchi_{\cquark 1}}\xspace}
\def\chictwo  {\ensuremath{\Pchi_{\cquark 2}}\xspace}
  \def\Y#1S{\ensuremath{\PUpsilon{(#1S)}}\xspace}

\def\chic  {\ensuremath{\Pchi_{c}}\xspace}

\def\proton      {\ensuremath{\Pp}\xspace}
\def\antiproton  {\ensuremath{\overline \proton}\xspace}

\def\Lbar {\ensuremath{\kern 0.1em\overline{\kern -0.1em\PLambda}}\xspace}

\def\BF         {{\ensuremath{\cal B}\xspace}}

\def\BR         {\BF}

\def\to                 {\ensuremath{\rightarrow}\xspace}

\def\AT#1     {\ensuremath{A_{\mathrm{T}}^{#1}}\xspace}           

\def\C#1      {\ensuremath{\mathcal{C}_{#1}}\xspace}                       
\def\Cp#1     {\ensuremath{\mathcal{C}_{#1}^{'}}\xspace}                    
\def\Ceff#1   {\ensuremath{\mathcal{C}_{#1}^{\mathrm{(eff)}}}\xspace}        
\def\Cpeff#1  {\ensuremath{\mathcal{C}_{#1}^{'\mathrm{(eff)}}}\xspace}       
\def\Ope#1    {\ensuremath{\mathcal{O}_{#1}}\xspace}                       
\def\Opep#1   {\ensuremath{\mathcal{O}_{#1}^{'}}\xspace}                    

\newcommand{\tev}{\ensuremath{\mathrm{\,Te\kern -0.1em V}}\xspace}
\newcommand{\gev}{\ensuremath{\mathrm{\,Ge\kern -0.1em V}}\xspace}
\newcommand{\mev}{\ensuremath{\mathrm{\,Me\kern -0.1em V}}\xspace}
\newcommand{\kev}{\ensuremath{\mathrm{\,ke\kern -0.1em V}}\xspace}
\newcommand{\ev}{\ensuremath{\mathrm{\,e\kern -0.1em V}}\xspace}
\newcommand{\gevc}{\ensuremath{{\mathrm{\,Ge\kern -0.1em V\!/}c}}\xspace}
\newcommand{\gevcnosp}{\ensuremath{{\mathrm{Ge\kern -0.1em V\!/}c}}\xspace}
\newcommand{\mevc}{\ensuremath{{\mathrm{\,Me\kern -0.1em V\!/}c}}\xspace}
\newcommand{\gevcc}{\ensuremath{{\mathrm{\,Ge\kern -0.1em V\!/}c^2}}\xspace}
\newcommand{\gevgevcccc}{\ensuremath{{\mathrm{\,Ge\kern -0.1em V^2\!/}c^4}}\xspace}
\newcommand{\mevcc}{\ensuremath{{\mathrm{\,Me\kern -0.1em V\!/}c^2}}\xspace}

\def\mum  {\ensuremath{\,\upmu\rm m}\xspace}

\def\invfb   {\ensuremath{\mbox{\,fb}^{-1}}\xspace}

\def\Xrad {\ensuremath{X_0}\xspace}

\newcommand{\stat}{\ensuremath{\mathrm{(stat)}}\xspace}
\newcommand{\syst}{\ensuremath{\mathrm{(syst)}}\xspace}

\newcommand{\chisq}{\ensuremath{\chi^2}\xspace}

\def\gsim{{~\raise.15em\hbox{$>$}\kern-.85em
          \lower.35em\hbox{$\sim$}~}\xspace}
\def\lsim{{~\raise.15em\hbox{$<$}\kern-.85em
          \lower.35em\hbox{$\sim$}~}\xspace}

\def\ptot       {\mbox{$p$}\xspace}
\def\pt         {\mbox{$p_{\rm T}$}\xspace}

\def\evtgen     {\mbox{\textsc{EvtGen}}\xspace}
\def\pythia     {\mbox{\textsc{Pythia}}\xspace}

\def\geant      {\mbox{\textsc{Geant4}}\xspace}

\def\photos     {\mbox{\textsc{Photos}}\xspace}

\def\tell1  {TELL1\xspace}
\def\ukl1   {UKL1\xspace}

\def\ptjpsi {\mbox{$p_{\rm T}^{\jpsi}$}\xspace}
\def\ptg {\mbox{$p_{\rm T}^{\g}$}\xspace}
\def\ptchic {\mbox{$p_{\rm T}^{\chic}$}\xspace}

\usepackage{cite} 
\usepackage{mciteplus}

\begin{document}

\renewcommand{\thefootnote}{\fnsymbol{footnote}}
\setcounter{footnote}{1}


\begin{titlepage}
\pagenumbering{roman}

\vspace*{-1.5cm}
\centerline{\large EUROPEAN ORGANIZATION FOR NUCLEAR RESEARCH (CERN)}
\vspace*{1.5cm}
\hspace*{-0.5cm}
\begin{tabular*}{\linewidth}{lc@{\extracolsep{\fill}}r}
\ifthenelse{\boolean{pdflatex}}
{\vspace*{-2.7cm}\mbox{\!\!\!\includegraphics[width=.14\textwidth]{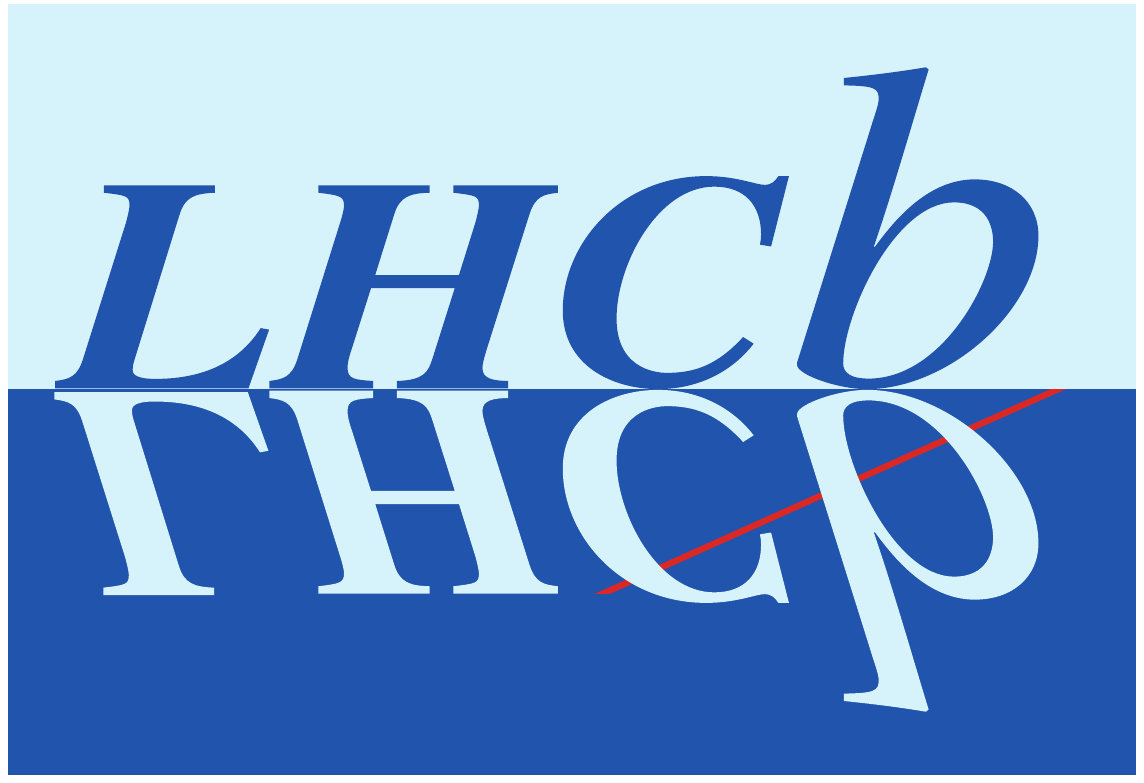}} & &}%
{\vspace*{-1.2cm}\mbox{\!\!\!\includegraphics[width=.12\textwidth]{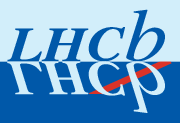}} & &}%
\\
 & & CERN-PH-EP-2013-114 \\  
 & & LHCb-PAPER-2013-028 \\  
 & & September 9, 2013 \\ 
 & & \\
\end{tabular*}

\vspace*{4.0cm}

{\bf\boldmath\huge
\begin{center}
Measurement of the relative rate of prompt \chiczero, \chicone and \chictwo production at $\sqrt{s}=7\tev$
\end{center}
}

\vspace*{2.0cm}

\begin{center}
The LHCb collaboration\footnote{Authors are listed on the following pages.}
\end{center}

\vspace{\fill}

\begin{abstract}
  \noindent
Prompt production of charmonium \chiczero, \chicone and \chictwo mesons is studied using proton-proton 
collisions at the LHC at a centre-of-mass energy of $\sqrt{s}=7\tev$. 
The $\chi_{c}$ mesons are identified through their decay to $\jpsi\gamma$, with \jpsi\to\mumu using
photons that converted in the detector. 
A data sample, corresponding to an integrated luminosity of 1.0\invfb collected by the LHCb 
detector, is used to measure the relative prompt production rate of \chicone and \chictwo 
in the rapidity range $2.0<y<4.5$ as a function of the \jpsi transverse momentum from 3 to 20 \gevc. 
First evidence for \chiczero meson production at a high-energy hadron collider is also presented.
\end{abstract}

\vspace*{1.5cm}

\begin{center}
  Submitted to JHEP.
\end{center}

\vspace{\fill}

{\footnotesize 
\centerline{\copyright~CERN on behalf of the \lhcb collaboration, license \href{http://creativecommons.org/licenses/by/3.0/}{CC-BY-3.0}.}}
\vspace*{2mm}

\end{titlepage}


\newpage
\setcounter{page}{2}
\mbox{~}
\newpage

\centerline{\large\bf LHCb collaboration}
\begin{flushleft}
\small
R.~Aaij$^{40}$, 
B.~Adeva$^{36}$, 
M.~Adinolfi$^{45}$, 
C.~Adrover$^{6}$, 
A.~Affolder$^{51}$, 
Z.~Ajaltouni$^{5}$, 
J.~Albrecht$^{9}$, 
F.~Alessio$^{37}$, 
M.~Alexander$^{50}$, 
S.~Ali$^{40}$
% ...

{\footnotesize \it
% ...
$ ^{5}$Clermont Universit\'{e}, Universit\'{e} Blaise Pascal, CNRS/IN2P3, LPC, Clermont-Ferrand, France\\
$ ^{6}$CPPM, Aix-Marseille Universit\'{e}, CNRS/IN2P3, Marseille, France\\
% ...
$ ^{9}$Fakult\"{a}t Physik, Technische Universit\"{a}t Dortmund, Dortmund, Germany\\
% ...
$ ^{36}$Universidad de Santiago de Compostela, Santiago de Compostela, Spain\\
$ ^{37}$European Organization for Nuclear Research (CERN), Geneva, Switzerland\\
% ...
$ ^{40}$Nikhef National Institute for Subatomic Physics, Amsterdam, The Netherlands\\
% ...
$ ^{45}$H.H. Wills Physics Laboratory, University of Bristol, Bristol, United Kingdom\\
% ...
$ ^{50}$School of Physics and Astronomy, University of Glasgow, Glasgow, United Kingdom\\
$ ^{51}$Oliver Lodge Laboratory, University of Liverpool, Liverpool, United Kingdom\\
% ...
$ ^{s}$Scuola Normale Superiore, Pisa, Italy\\
}
\end{flushleft}

\cleardoublepage


\renewcommand{\thefootnote}{\arabic{footnote}}
\setcounter{footnote}{0}



\pagestyle{plain} 
\setcounter{page}{1}
\pagenumbering{arabic}

%

\section{Introduction}
\label{sec:Introduction}
The study of charmonium production provides an important test of the underlying mechanisms described by quantum chromodynamics (QCD).
In $pp$ collisions charmonia can be produced directly, or indirectly via the decay of higher excited states (feed-down)
or via the decay of $b$~hadrons. The first two are referred to as prompt production. The mechanism
for the production of the prompt component is not yet fully understood, and none of the
available models adequately predicts both the transverse momentum spectrum and the  polarization of the promptly 
produced charmonium states~\cite{Brambilla2011}.

At the \lhc, \ccbar pairs are expected to be produced at leading order (LO) through gluon-gluon interactions,
followed by the formation of bound charmonium states. The production of the \ccbar pair is described by perturbative QCD
while non-perturbative QCD is needed for the description of the evolution of the \ccbar pair to the bound state.
Several models have been developed for the non-perturbative part, such as 
the Colour Singlet (CS) model~\cite{CSMLikhoded,CSMBerger,CSMBaier} and the non-relativistic QCD (NRQCD) model~\cite{NRQCD}.
The CS model  assumes the \ccbar pair is created in a hard scattering reaction as a colour singlet with the same quantum numbers 
as the final charmonium state. 
The NRQCD model  includes,  in addition to the colour singlet mechanism, the production of \ccbar pairs as colour octets (CO) 
(in this case the CO state evolves to the final charmonium state via soft gluon  emission).
These two models predict different ratios of the \chictwo to \chicone production cross-sections.

The study of the production of \chic states is also important since these resonances give a substantial feed-down contribution to
prompt \jpsi production~\cite{LHCb-PAPER-2011-030} through their radiative decay $\chic\to\jpsi\gamma$ and can have a significant impact on the
\jpsi polarization measurement~\cite{LHCb-PAPER-2013-008}. 
Measurements of \chicone and \chictwo production cross-section for various particle beams and energies have been reported
in Refs.~\cite{WA11chic,HERABchic,CDFchic,CMSchic,LHCb-PAPER-2011-019}. 

In this paper we report a measurement of the ratio of  prompt \chictwo to \chicone production cross-sections $\sigma(pp\to\chictwo X)/\sigma(pp\to\chicone X)$
at a centre-of-mass energy of $\sqrt{s}=7\tev$ in the rapidity range $2.0<y<4.5$
as a function of the \jpsi transverse momentum (\pt) from 3 to 20\gevc. The data sample corresponds 
to an integrated luminosity of 1.0\,fb$^{-1}$ collected during 2011 by the LHCb detector.
The radiative decay $\chic\to\jpsi\gamma$ is used, where the \jpsi is reconstructed in the dimuon final state and only photons 
that convert in the detector material are used. The converted photons are reconstructed using \ep and \en tracks, which allows a 
clean separation of the \chicone and \chictwo peaks, due to a better 
energy resolution of converted photons than for those that are identified with the calorimeter (referred to as calorimetric photons in
the following). 

The measurement performed by LHCb using calorimetric  photons with 2010 data~\cite{LHCb-PAPER-2011-019}
 was limited by the fact that the two \chic peaks were not well separated.
The measurements with calorimetric~\cite{LHCb-PAPER-2011-019} and converted (as presented in this study) photons 
are largely uncorrelated since the photon reconstruction is based on different subdetectors.
Furthermore, this is the first measurement using converted photons in LHCb.
The \chiczero state has been previously observed in ${\ensuremath{\Pp}}\antiproton$ collisions at threshold~\cite{E835}, 
but this letter reports the first evidence at high-energy  hadron colliders. Its production rate relative to that of the \chictwo is also reported.

\section{The LHCb detector and dataset}
The \lhcb detector~\cite{Alves:2008zz} is a single-arm forward
spectrometer covering the \mbox{pseudorapidity} range $2<\eta <5$,
designed for the study of particles containing \bquark or \cquark
quarks. The detector includes a high precision tracking system
consisting of a silicon-strip vertex detector (VELO) surrounding the $pp$
interaction region, a large-area silicon-strip detector located
upstream of a dipole magnet with a bending power of about
$4{\rm\,Tm}$, and three stations of silicon-strip detectors and straw
drift tubes placed downstream. 
The combined tracking system provides a momentum measurement with
relative uncertainty that varies from 0.4\% at 5\gevc to 0.6\% at 100\gevc,
and impact parameter resolution of 20\mum for
tracks with high transverse momentum. 
Charged hadrons are identified
using two ring-imaging Cherenkov detectors. 
Electron and hadron candidates are identified by a calorimeter system consisting of
scintillating-pad (SPD) and preshower detectors, an electromagnetic
calorimeter (ECAL) and a hadronic calorimeter. 
The SPD and preshower are designed to distinguish between signals from photons and electrons.
The ECAL is constructed from scintillating tiles interleaved with lead tiles.
The reconstruction of converted photons that are used in this analysis is described in Sec.~\ref{sec:Selection}.
Muons are identified by a system composed of alternating layers of iron and multiwire
proportional chambers.
The total radiation length before the first tracking station is about 0.25\Xrad~\cite{Alves:2008zz}.

The LHCb coordinate system is defined to be right-handed with its origin at the nominal interaction point, the
$z$ axis aligned along the beam line towards the magnet and the
$y$ axis pointing upwards. The magnetic field is oriented along the $y$ axis.

The trigger~\cite{LHCb-DP-2012-004} consists of a
hardware stage, based on information from the calorimeter and muon
systems, followed by a software stage, which applies a full event
reconstruction. 
Candidate events used in this analysis are first required to pass a hardware trigger,
which selects muons with  $\pt>1.48\gevc$ or dimuon candidates with a product of their \pt larger than
$1.68~(\gevcnosp)^2$.
In the subsequent software trigger, both muons are required to have
$\pt>0.5\gevc$, total momentum $\ptot>6\gevc$, and dimuon invariant mass greater than $2.5\gevcc$.

In the simulation, $pp$ collisions are generated using
\pythia~6.4~\cite{Sjostrand:2006za} with a specific \lhcb
configuration~\cite{LHCb-PROC-2010-056}. The NRQCD matrix elements are used in \pythia~6.4. Decays of hadronic particles
are described by \evtgen~\cite{Lange:2001uf}, in which final state
radiation is generated using \photos~\cite{Golonka:2005pn}. The
interaction of the generated particles with the detector and its
response are implemented using the \geant
toolkit~\cite{Allison:2006ve, *Agostinelli:2002hh} as described in
Ref.~\cite{LHCb-PROC-2011-006}. 
The simulated samples consist of events in which at least one $\jpsi\to\mumu$ decay takes place.
In a first sample used for background studies there is no constraint on the \jpsi production mechanism.
In the second sample used for the estimation of signal efficiencies the 
\jpsi is required to originate from a \chic meson.

\section{Event reconstruction and selection}
\label{sec:Selection}
Photons that convert in the detector material are reconstructed from a pair of oppositely charged electron candidates. 
Since photons that have converted in the VELO have lower acceptance and worse energy resolution,
only $\gamma\to e^+e^-$ candidates without VELO hits are considered. This selection strongly favours
conversions that occur between the downstream end of the VELO and the first tracking station upstream of the
magnet.

Candidate $e^+e^-$ pairs are required to be within the ECAL acceptance and produce electromagnetic clusters that have compatible 
$y$ positions.
A bremsstrahlung correction is applied to each electron track: any photon whose position in the ECAL is compatible with a straight line extrapolation 
of the electron track from the first tracking stations is selected and its energy is added to the electron energy from the reconstructed track. 
If the same bremsstrahlung candidate is found for both the \ep and the \en of the pair, the photon energy is added randomly to one of the tracks.
The \ep and \en tracks (corrected for bremsstrahlung) are then extrapolated backward in order to determine the conversion point and a vertex fit 
is performed  to reconstruct the photon.
The photon's invariant mass is required to be less than 100\mevcc.
Combinatorial background is suppressed by applying a cut on the \epem invariant mass ($M_{\epem}$) such that 
 $M_{\epem} < 0.04\times z_{\rm vtx} + 20\,\mevcc$ where $z_{\rm vtx}$ is the $z$ coordinate of the conversion in mm.
Converted photons are required to have  transverse momentum (\ptg) greater than 0.6\gevc.

The \jpsi candidate is reconstructed in its decay to $\mumu$. Each track must be identified as a muon with $\pt>0.65\gevc$, $\ptot>6\gevc$ 
and a track fit $\chisq/\rm{ndf}$ smaller than 5, where ndf is the number of degrees of freedom.
The two muons must originate from a common vertex with vertex fit $\chisq_{\rm{vtx}}/\rm{ndf}$ smaller than 20. 
In addition the \mumu invariant mass is required to be in the range 3058--3138\mevcc.

\begin{figure}[tb]
  \begin{center}
    \includegraphics[width=0.7\linewidth]{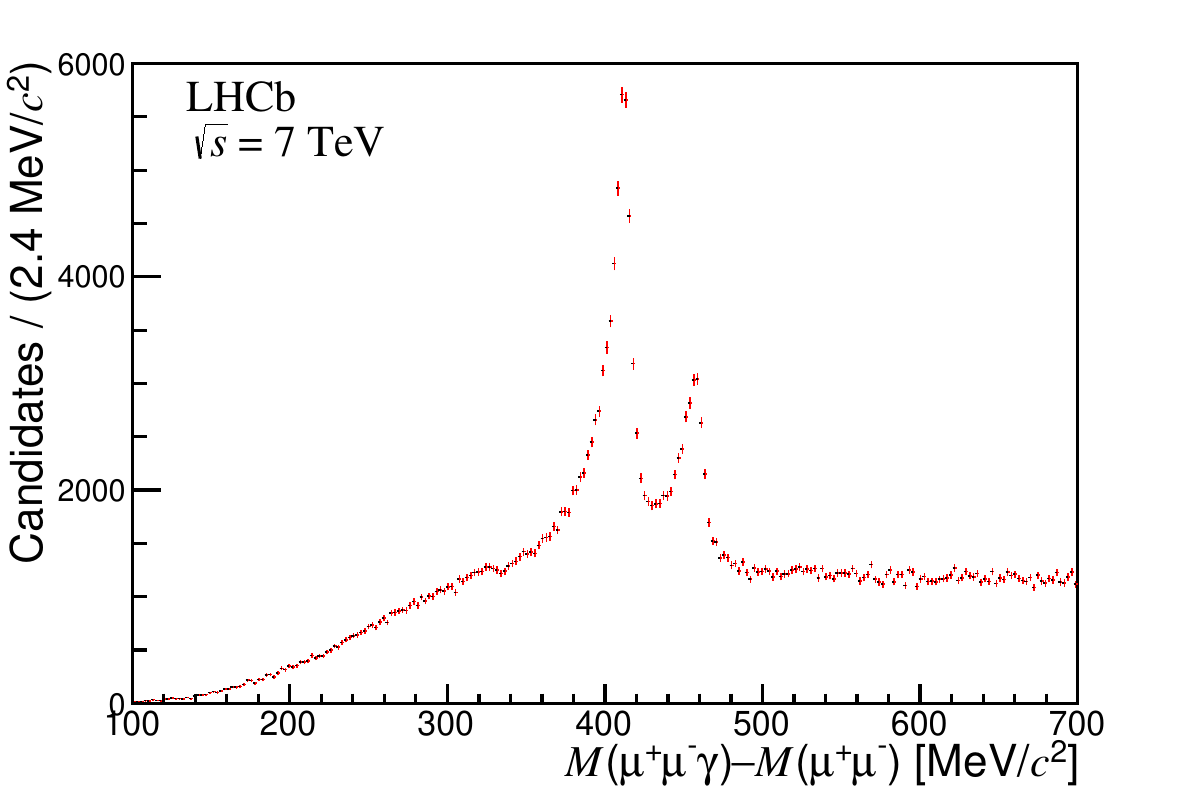} 
\vspace{-.5cm}
  \end{center}
  \caption{
    \small Distribution of the mass difference $\Delta M\equiv M(\mumu\g)-M(\mumu)$ for \chic candidates with $3<\ptjpsi<20\gevc$. 
}  \label{fig:dMall}
\end{figure}

The \jpsi and \g candidates are associated with the primary vertex (PV) to which they have the smallest impact parameter. 
These \jpsi and photon candidates are combined to form a \chic candidate.
Loose requirements are applied in order to reject combinatorial background and poorly reconstructed candidates using the following variables:
the difference in $z$-positions of the primary vertices associated with the \jpsi and \g,
the \chisq of the \chic candidate vertex fit  and the difference between the \chisq of the PV reconstructed with and without the \chic candidate.
These cuts remove about $20\%$ of the background and $5\%$ of the signal.
Contributions from $b\to\chic X$ are suppressed by requiring that the \chic decay time is smaller than 0.15~ps.
This removes about 85\% of non-prompt events and 0.5\% of the prompt \chic signal.
Figure~\ref{fig:dMall} shows the distribution of the difference in the invariant masses of the \chic  and \jpsi selected candidates 
$\Delta M\equiv M(\mumu\g)-M(\mumu)$ for candidates with  \jpsi transverse momentum (\ptjpsi) in the range 3--20\gevc.

\section{Determination of the ratio of cross-sections}
\label{sec:signal}
The production cross-section ratio of the \chictwo and \chicone mesons is measured in 
ten \ptjpsi bins of different width (the bin limits are given in Table~\ref{tbl:systematics}) with
\begin{equation}
\frac{\sigma(\chictwo)}{\sigma(\chicone)}=\frac{N_{\chictwo}}{N_{\chicone}}\frac{\varepsilon_{\chicone}}{\varepsilon_{\chictwo}}\frac{\BR(\chicone\to\jpsi\g)}{\BR(\chictwo\to\jpsi\g)},
\label{eq:rxs}
\end{equation}
where $\sigma(\chi_{cJ})$ is the prompt $\chi_{cJ}$ production cross-section, $N_{\chi_{cJ}}$ is the prompt $\chi_{cJ}$ yield ($J=1,2$), and
$\BR(\chicone\to\jpsi\g)=(34.4\pm1.5)\%$ and 
$\BR(\chictwo\to\jpsi\g)=(19.5\pm0.8)\%$~\cite{PDG2012} 
are the known branching fractions. 
The efficiency ratio is expressed as
\begin{equation}
\frac{\varepsilon_{\chicone}}{\varepsilon_{\chictwo}}=\frac{\varepsilon^{\jpsi}_{\chicone}}{\varepsilon^{\jpsi}_{\chictwo}}\frac{\varepsilon^{\g}_{\chicone}}{\varepsilon^{\g}_{\chictwo}},
\label{eq:reff}
\end{equation}
where $\varepsilon^{\jpsi}_{\chi_{cJ}}$ is the efficiency to trigger, detect, reconstruct and select a \jpsi from a $\chi_{cJ}$ decay and $\varepsilon^{\g}_{\chi_{cJ}}$ 
is the efficiency to detect, reconstruct and select a photon from a $\chi_{cJ}$ decay once the \jpsi has been selected and then to select the $\chi_{cJ}$ meson. 
The efficiency $\varepsilon^{\g}_{\chi_{cJ}}$ 
includes the probability for a photon to convert upstream of the first tracking station (about $20\%$). 

The ratio $\sigma(\chiczero)/\sigma(\chictwo)$ is also measured with appropriate substitutions in Eqs.~\ref{eq:rxs} and~\ref{eq:reff} and 
using the known value $\BR(\chiczero\to\jpsi\g)=(1.17\pm0.08)\%$~\cite{PDG2012}.
Due to this small branching fraction, the number of reconstructed \chiczero mesons is also small and therefore the ratio of production cross-sections 
is only measured in one wide \ptjpsi bin, 4--20~\gevc. The  \chiczero cross-section is measured relative to the \chictwo cross-section rather than to
the \chicone cross-section because the \pt dependence is expected to be similar inside this \pt range for \chiczero and \chictwo~\cite{Likhoded_chic}.
\subsection{Background studies}
\label{sec:bkg}
There are two sources of background: a peaking component from non-prompt \chic (from $b$-hadron decays) production and a non-peaking 
combinatorial contribution.

The peaking background is estimated by fitting the decay time distribution of the \chic candidates with decay time larger than 0.3~ps 
with an exponential shape and extrapolating into the signal region ($0-0.15$~ps). 
The combinatorial background from $b$-hadron decays lying under the peak is evaluated using the 
lower or upper mass sidebands. The two estimates agree and the average is used to subtract its contribution. 
The simulation predicts that \chic mesons from $b$-hadron decays tend to be more energetic than  prompt \chic mesons.
The fraction of peaking background is  therefore estimated in two regions of \ptjpsi, below and above 9~\gevc, and the maximum deviation from the 
mean value inside each range 
(as predicted by simulation) is taken as a systematic uncertainty. For the \chicone meson 
the remaining peaking background is $(0.9\pm0.3)\%$ 
of the signal for \ptjpsi below 9\gevc and $(1.8\pm0.4)\%$ above this value.
As expected~\cite{PDG2012,LHCb-PAPER-2013-024} the number of non-prompt \chictwo candidates is smaller.   
The relative yield of non-prompt \chictwo and \chicone mesons
 is obtained from a fit to the $\Delta M$ distribution of the events 
rejected by the cut on the $\chic$ decay time (using the method described in Sec.~\ref{sec:fit}). 
The ratio of branching fractions is determined to be 
\begin{equation*}
\frac{\BR(b\to\chictwo)\times \BR(\chictwo\to\jpsi\g)}{\BR(b\to\chicone)\times \BR(\chicone\to\jpsi\g)}=0.184\pm0.025\,\stat\pm0.015\,\syst,
\end{equation*} 
where the systematic uncertainty is obtained by varying the fit function parameters. 
The remaining number of non-prompt \chictwo candidates is then determined as the number of remaining non-prompt \chicone mesons 
multiplied by this ratio of branching fractions.
For the \chiczero peak it is not possible to estimate the non-prompt contribution from the data but this is expected to be at most $2\%$. This assertion
is based on the similar values for $\BR(b\to\chicone X)$ and $\BR(b\to\chiczero X)$~\cite{PDG2012}  and the small contamination
of $b\to\chicone X$ decays as shown above.
Another peaking background arises from the decay of prompt \psitwos to a \chic meson. 
According to simulation and cross-section measurements~\cite{LHCb-PAPER-2011-045} this background can be safely neglected.

The shape of the combinatorial background is estimated using the selected data sample by generating ``fake photons'' to mimic the candidate photon spectra in data. 
For each $\chic\to\jpsi\gamma$ candidate, two fake photons are generated: one where the photon energy is set equal to twice the \en energy, 
and a second where twice the \ep energy is used. 
In this way, a spread of fake photon energies are produced, all with the same angular distribution as the candidate photons in the data. 
Each of these photons is then combined with the \jpsi candidate to form the fake \chic candidate.
The contribution from the \chic signal region is normalized to the estimated background contribution in the same invariant mass region 
(this procedure converges with few iterations). 
The procedure was tested on simulated events and reproduces the $\Delta M$ distribution of the combinatorial background in the region of the 
\chicone and \chictwo signal peaks.
\subsection{Efficiency corrections}
\label{sec:effic}
The ratio of the overall efficiencies for the detection of \jpsi mesons originating from the decay of a \chicone meson compared to a \chictwo meson,
$\varepsilon^{\jpsi}_{\chicone}/\varepsilon^{\jpsi}_{\chictwo}$, is estimated from simulation and is compatible with unity for all \ptjpsi bins.

Since the kinetic energy released  in the \chicone decay ($Q$-value) is smaller 
than that of the \chictwo decay, the photon \pt spectrum differs for the two decays. 
As a result, the photon \pt requirement ($\ptg>0.6$\gevc) has a lower efficiency for the \chicone decay. 
Moreover the reconstruction efficiency of the converted photon decreases as the photon \pt decreases.
This is due to the fact that low energy electrons escape the detector before reaching the calorimeter and are therefore not identified as electrons.
Thus, the efficiency ratio is expected to be smaller than unity. 
The value obtained from simulation is  $\varepsilon^{\g}_{\chicone}/\varepsilon^{\g}_{\chictwo}=0.95\pm0.01$ and shows no significant 
dependence on $\ptjpsi$.

\begin{figure}[tb]
  \begin{center}
    \begin{tabular}{cc}
     \hspace{-.7cm}
\includegraphics[width=0.52\linewidth]{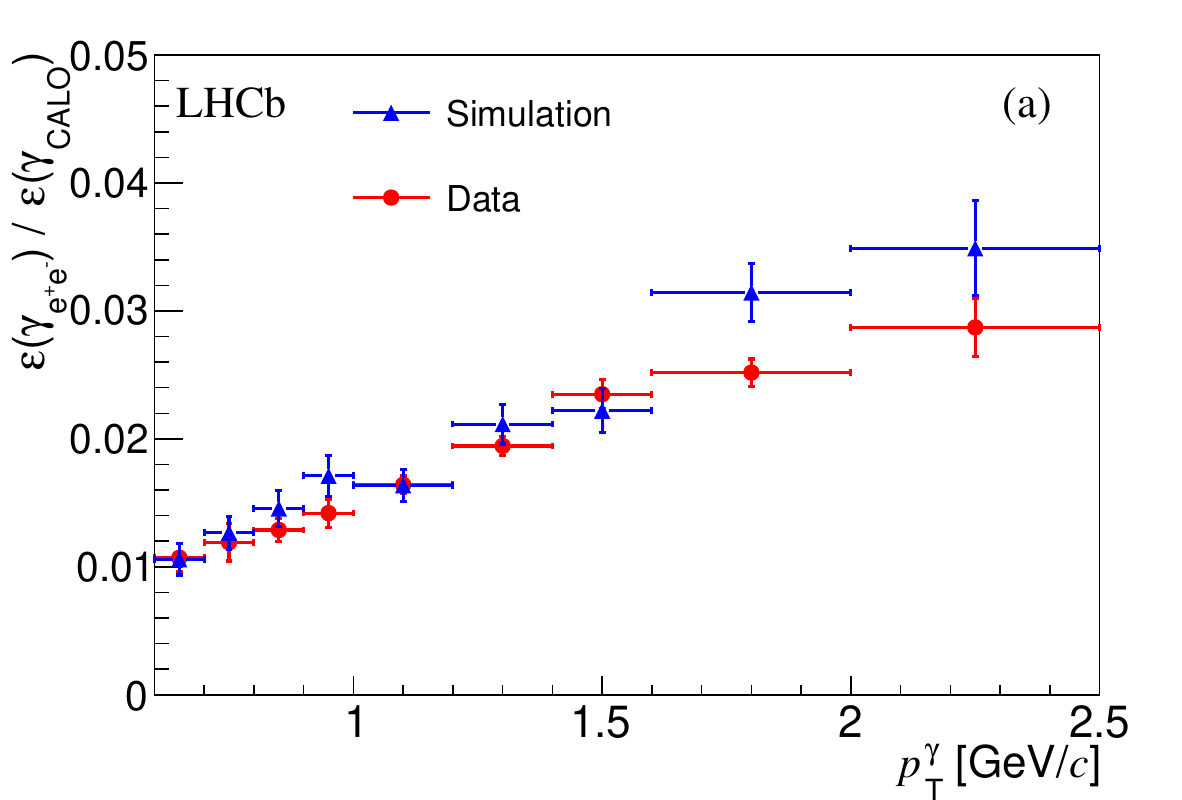} &
      \hspace{-.7cm}   
\vspace{-1mm}\includegraphics[width=0.52\linewidth]{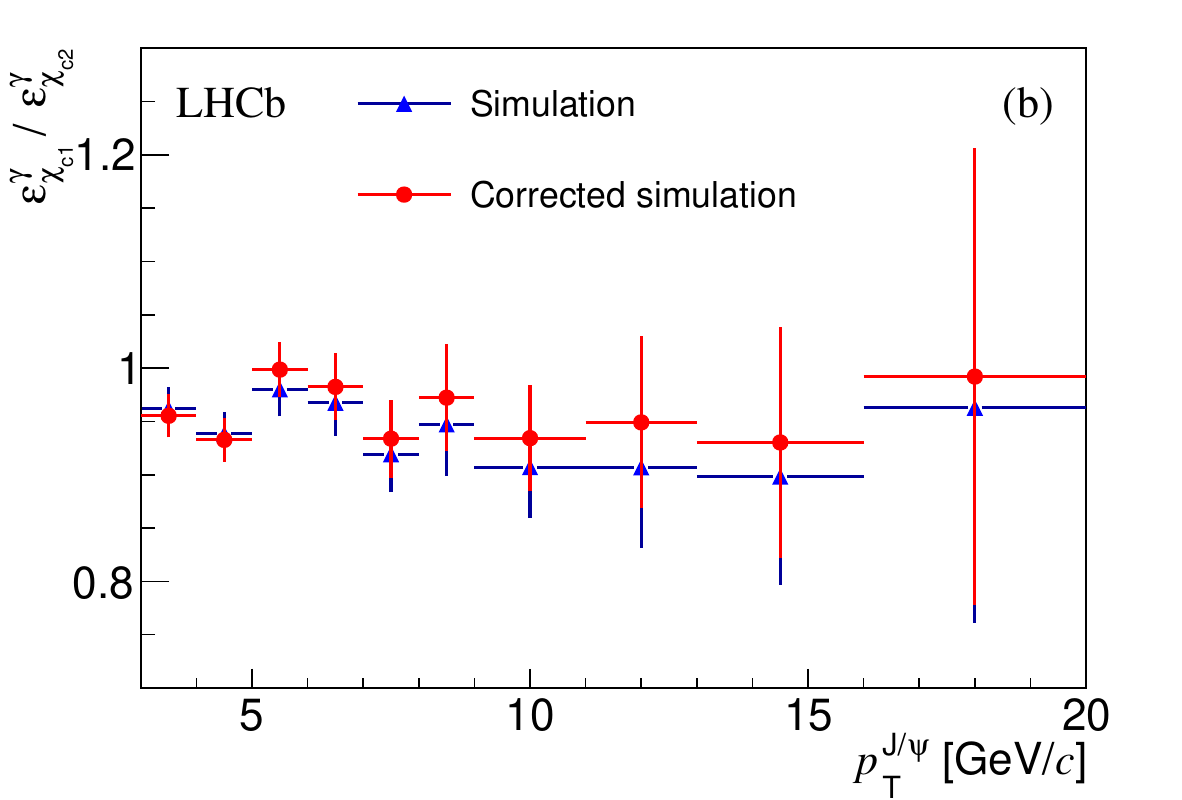} \\
      \end{tabular}
   \vspace{-.5cm}
  \end{center}
  \caption{\small 
(a) Efficiency of converted photon reconstruction and selection relative to the calorimetric photon efficiency for data (red circles) 
and simulated events (blue triangles) as a function of \ptg.
(b) Ratio of photon efficiencies $\varepsilon^{\g}_{\chicone}/\varepsilon^{\g}_{\chictwo}$ as a function of \ptjpsi 
from simulation (blue triangles) and after correcting the simulation
for the converted photon efficiency measured in data (red circles) taken from plot (a).
}
  \label{fig:efficphot}
\end{figure}

The conversion probability and total efficiency for converted photons is cross-checked using \piz mesons, 
reconstructed either with two calorimetric photons or with one calorimetric photon and one converted photon. The ratio of efficiencies of converted photons 
to calorimetric photons is measured in data and simulation as a function of \ptg and is shown in Fig.~\ref{fig:efficphot}(a).
The total efficiency for calorimetric photons is described well by simulation~\cite{LHCb-PAPER-2013-024} therefore these measurements give a direct comparison of 
the converted photon efficiency in data and simulation.
The efficiency with which converted photons are reconstructed in simulation is consistent with data (within about $15\%$).
 The results obtained from this  study are used to correct the simulation. The corrected
$\varepsilon^{\g}_{\chicone}/\varepsilon^{\g}_{\chictwo}$ ratio is shown as a function of \ptjpsi in Fig.~\ref{fig:efficphot}(b). 
This ratio is still compatible with a constant: 
$\varepsilon^{\g}_{\chicone}/\varepsilon^{\g}_{\chictwo}=0.96\pm0.01$. 

For the \chiczero to \chictwo ratio the corrected efficiency ratio is $\varepsilon_{\chictwo}/\varepsilon_{\chiczero}=1.69\pm0.18$. 
The departure from unity is due to the different $Q$-values of the two decays, as discussed above.
\subsection{Determination of the yield ratios}
\label{sec:fit}
The $\Delta M$ spectrum is fitted to determine the signal yields.
The \chicone and \chictwo signal peaks are each parametrized with a double-sided Crystal Ball (CB) function~\cite{Skwarnicki:1986xj}
\begin{eqnarray}
\label{eq:CB}
f_i(x)\propto & {\rm exp}(-\frac{1}{2}(\frac{x-\Delta M_i}{\sigma_i})^2) & ~\mbox{ for }~ -\alpha_L<\frac{x-\Delta M_i}{\sigma_i}<\alpha_R  \nonumber
\\
f_i(x)\propto & \frac{(n_L/\alpha_L)^{n_L} {\rm exp}(-\frac{1}{2}\alpha_L^2)}{(n_L/\alpha_L-\alpha_L-(x-\Delta M_i)/\sigma_i)^{n_L}} & ~\mbox{ for }~ \frac{x-\Delta M_i}{\sigma_i}<-\alpha_L
\\
f_i(x)\propto & \frac{(n_R/\alpha_R)^{n_R} {\rm exp}(-\frac{1}{2}\alpha_R^2)}{(n_R/\alpha_R-\alpha_R+(x-\Delta M_i)/\sigma_i)^{n_R}} & ~\mbox{ for }~ \frac{x-\Delta M_i}{\sigma_i}>\alpha_R \nonumber ,
\end{eqnarray}
where  the index $i=1$ (2) refers to the \chicone (\chictwo) CB function.
The left tail accounts for events with unobserved bremsstrahlung photon(s) while
the right tail accounts for events reconstructed with background photons. Simulation shows that the same $\alpha$ and $n$ parameters can be used
for both the \chicone and \chictwo peaks and that the \chictwo mass resolution, $\sigma_2$, is $10\%$ larger than the \chicone mass resolution, $\sigma_1$. 
These constraints are used in all the fits.
A \chiczero contribution is also included and is modelled by the convolution of a CB and a Breit-Wigner distribution 
with the width set to the \chiczero natural width ($10.4\pm0.6$\mevcc~\cite{PDG2012}) and  with the peak position fixed from simulation. 
For the $\chiczero$ CB shape, the same tail parameters are used as for the \chicone and \chictwo CB functions.

The full data sample ($3<\ptjpsi<20$\gevc) after background subtraction is fitted with the sum of these three functions. 
The peak positions $\Delta M_1$ and $\Delta M_2$, the \chicone resolution $\sigma_1$ 
and the CB $n$ parameters 
obtained from this fit are then used for the individual fits in each \ptjpsi bin.
The same fit is performed on simulated \chic events (without background) and the value of the $n$ parameter is found compatible with the
data for the left tail while slightly smaller for the right tail. These values are used when studying systematic effects. The \chic mass 
resolution is also found to be significantly smaller in simulation due to better energy resolution in the reconstruction of converted photons.

\begin{figure}[tb]
  \begin{center}
\begin{tabular}{cc}
    \hspace*{-.85cm}
  \includegraphics[width=0.52\linewidth]{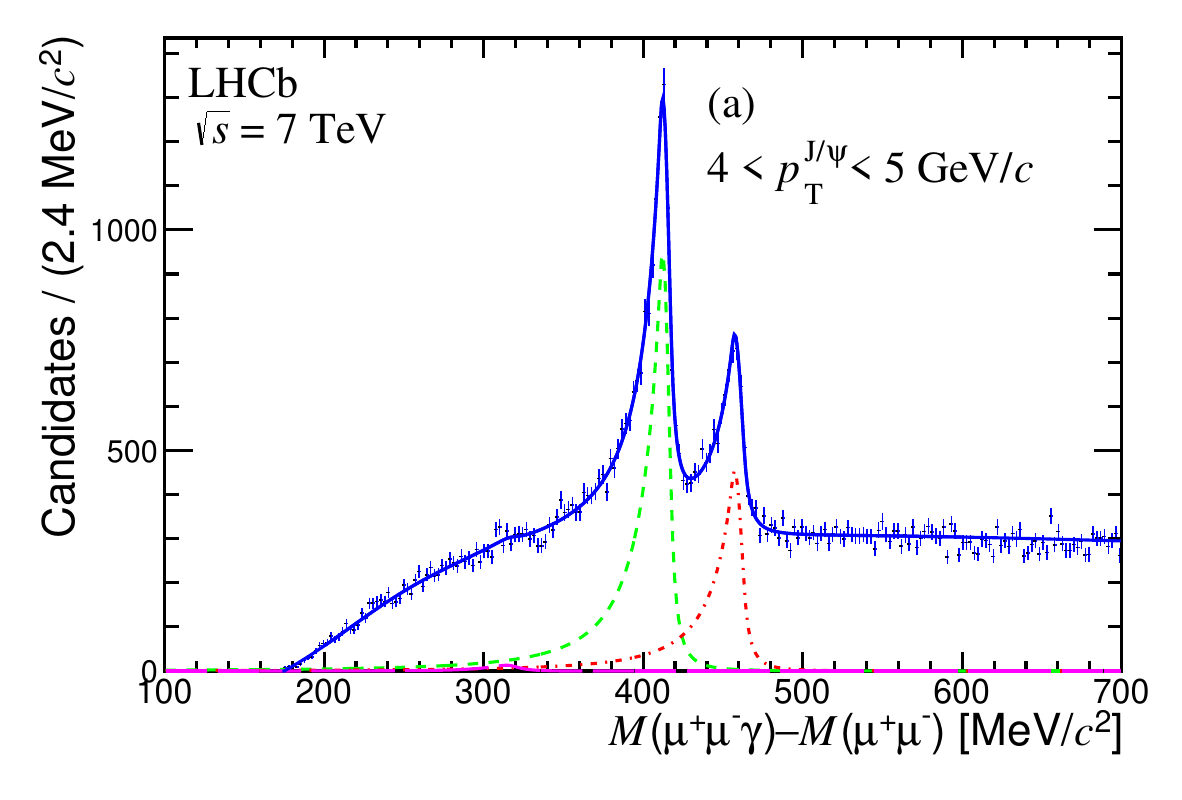} &
    \hspace*{-.65cm}
\includegraphics[width=0.52\linewidth]{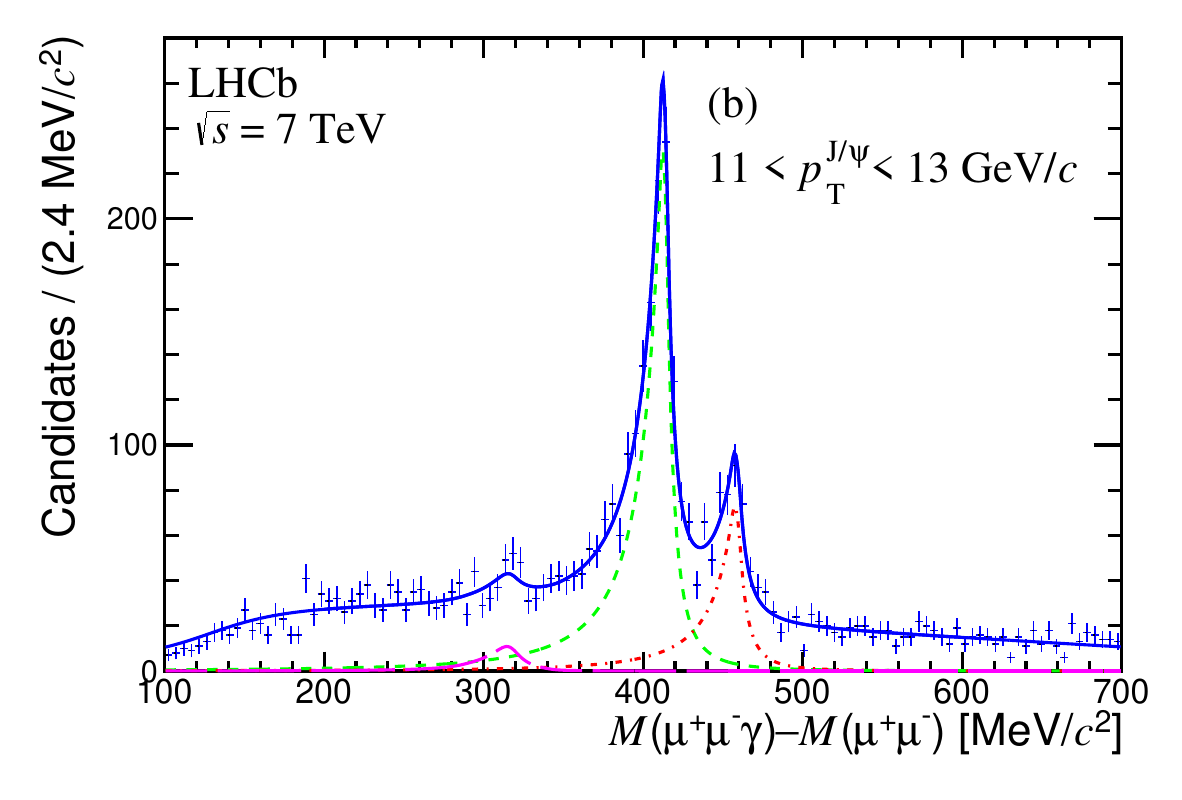} \\
   \vspace*{-1.0cm}
\end{tabular}
  \end{center}
  \caption{
    \small Distribution of $\Delta M=M(\mumu\g)-M(\mumu)$ for \ptjpsi in the range (a) 4--5~\gevc and (b) 11--13~\gevc. 
The results of the fit
are also shown, with  the total fitted function (blue solid curve), the \chicone signal (green dashed curve), the \chictwo signal 
(red dot-dashed curve) and the \chiczero signal (purple long-dashed curve).
}  \label{fig:dM}
\end{figure}

For each \ptjpsi bin the combinatorial background shape is determined using the candidates reconstructed with the  fake photons. 
The $\Delta M$ distribution of these
candidates is fitted with an empirical function
\begin{equation}
f_{\rm bkg}(\Delta M)\propto {\arctan}\biggl(\frac{\Delta M -m_0}{c}\biggr)+b\biggl(\frac{\Delta M}{m_0}-1\biggr)+a,
\end{equation} 
where $m_0$, $a$, $b$ and $c$ are free parameters.
This function is then used to parametrize the combinatorial background with all parameters fixed except for the normalization.
In total there are six free parameters for each fit: the CB function $\alpha$ parameters (left and right tails), the height of the \chicone and \chiczero 
peaks, the ratio of \chictwo to \chicone heights and the background normalization.
Figure~\ref{fig:dM} shows the $\Delta M$ distribution and the fit results
  for two ranges:  $4<\ptjpsi<5$~\gevc and $11<\ptjpsi<13$~\gevc. 

\begin{figure}[tb]
  \begin{center}
  \begin{tabular}{cc}
        \hspace*{-.5cm}\includegraphics[width=0.52\linewidth]{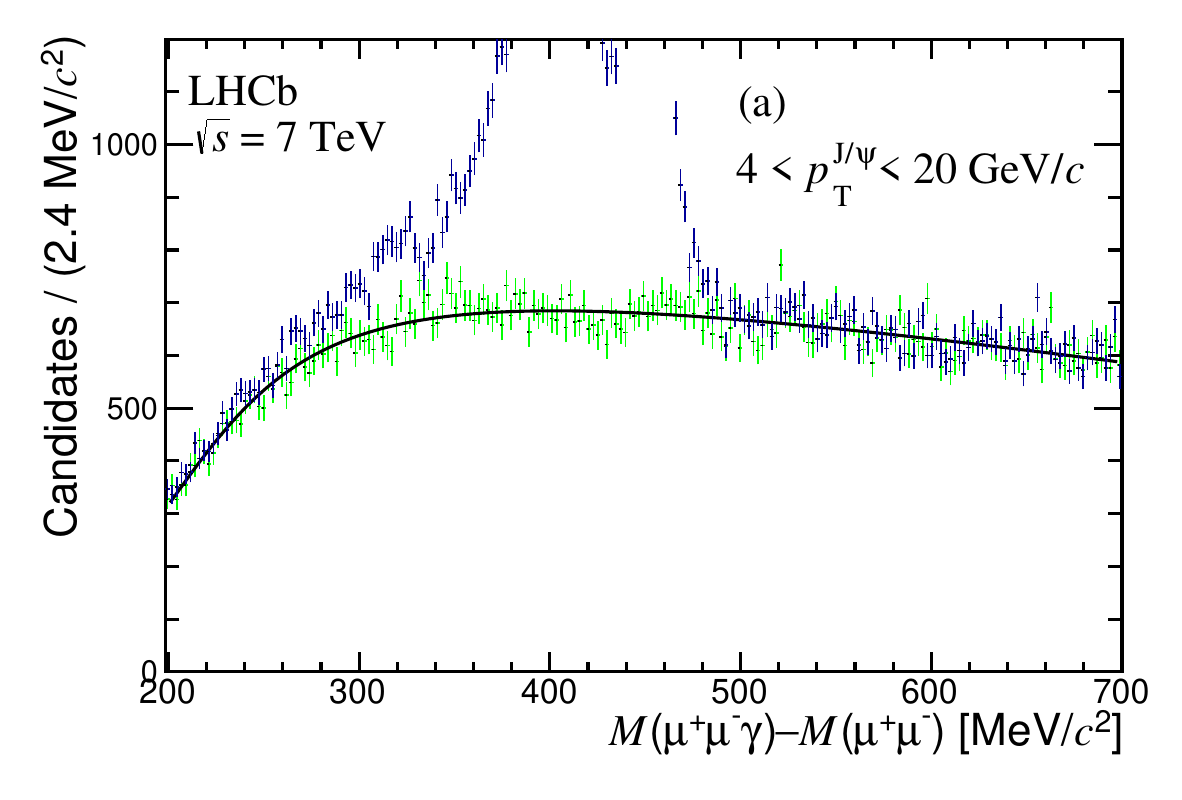} &
        \hspace*{-.65cm}\includegraphics[width=0.52\linewidth]{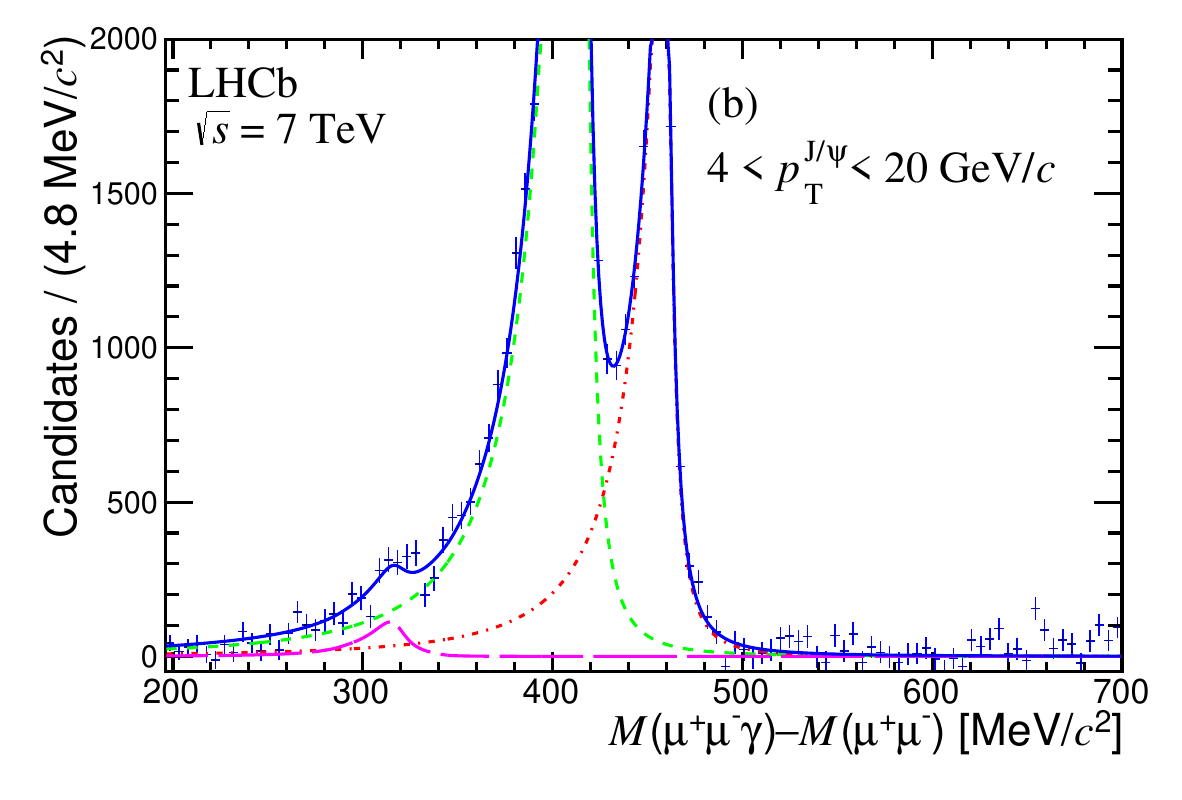} \\
    \vspace*{-1.0cm}
    \end{tabular}
  \end{center}
  \caption{
    \small Distribution of $\Delta M=M(\mumu\g)-M(\mumu)$ (blue histogram) for $4<\ptjpsi<20\gevc$. (a) The background estimated using fake photons (green) is superimposed 
on the  $\Delta M$  distribution, together with the function used to parametrize it (black solid line).
(b) The same $\Delta M$ distribution after background subtraction (using the shape shown in (a) and its fitted normalization):
total fitted function (blue solid curve), 
 \chicone signal (green dashed curve), \chictwo signal 
(red dot-dashed curve) and \chiczero signal (purple long-dashed curve).
\label{fig:chic0} 
}
\end{figure}

The \chiczero yield is not significant in the individual bins and is therefore only measured over the  integrated range $4<\ptjpsi<20\gevc$.
The region 3--4~\gevc is excluded because for this particular \ptjpsi bin the background is high  and not well modelled below 300\mevcc,
close to the \chiczero peak.
Figure~\ref{fig:chic0} (a) shows the total $\Delta M$ distribution superimposed with
the background estimate using the fake photons and the fit to this background distribution. The \chiczero contribution is visible just above $300\mevcc$.
Figure~\ref{fig:chic0} (b) shows the result of the fit for $4<\ptjpsi<20\gevc$ after background subtraction.

\section{Systematic uncertainties}
\label{sec:syst}
The fit is performed for each \ptjpsi bin as explained in Sec.~\ref{sec:signal}. The \chicone and \chictwo peak positions, the CB width and 
the left and right tail $n$ parameters are fixed to those found in the fit to the whole dataset.
In order to assess the stability, the fit is also performed with all parameters left free except for the peak positions or using the $n$ parameters
obtained with simulated events.
The fit is also repeated in a smaller range ($\Delta M>290\mevcc$) in order to assess the uncertainty coming from
 the imperfect modelling of the background at small $\Delta M$.
It is also repeated on the distribution with the background subtracted.
The largest variation from these alternative fits is taken as a systematic uncertainty. 
The fit quality is usually good (the $p$-values of the fits are greater than $1\%$) 
except for the first \ptjpsi bin where the background is not  well modelled for low $\Delta M$. However the ratio of $\chictwo$ and $\chicone$ yields 
is stable, indicating it is relatively insensitive to the modelling in this low $\Delta M$ region. 
For the \chiczero yield this systematic uncertainty is $20\%$ and is dominated by the variation of the $n_L$ parameter.
This large uncertainty is incurred because the \chiczero lies in the low mass tail of the \chicone mass spectrum, and is sensitive to the modelling of the 
\chicone signal shape.

The bias due to the fitting procedure is studied using simulated events.
This study indicates a bias of $(-4.8\pm1.8)\%$ and $(-2.4\pm2.0)\%$ for the first and second
 \ptjpsi bins, respectively, and therefore the data
are corrected for these biases. The other bins show no significant bias within the $3\%$ uncertainty of the test. 
Conservatively, a systematic uncertainty of $3\%$ is assigned to all bins.

Imperfect modelling of the combinatorial background may introduce a  bias. This is studied with simulated events by comparing the results
obtained using the $\Delta M$ distribution of true background events and the distribution of the background estimated with the fake photons.
The bias is found to be within $1\%$, which is assigned as a systematic uncertainty to all the bins. 
For the \chiczero yield the impact of an imperfect modelling of the background can be absorbed in the variation of the $n_L$ parameter of the \chicone 
CB function.
This is therefore already accounted for in the fit systematic uncertainty.

The peaking background (\chic from $b$ hadrons) is estimated in Sec.~\ref{sec:bkg} and is subtracted from the number of \chicone candidates:
$(0.9\pm0.3)\%$ for \ptjpsi below 9~\gevc and $(1.8\pm0.4)\%$ above. The number of \chictwo candidates is $0.18\pm0.03$ times the number of \chicone 
candidates (see Sec.~\ref{sec:bkg}).
The ratio of prompt \chic mesons is corrected for this background and a systematic uncertainty of $0.3\%$ ($0.4\%$) is assigned for the \ptjpsi bins below
(above) 9~\gevc. No peaking background correction is applied for the ratio of \chiczero to \chictwo yields. This correction is estimated
to be at most $2\%$  (see Sec.~\ref{sec:bkg}) which is taken as the systematic uncertainty.

The photon efficiency is discussed in Sec.~\ref{sec:effic}: the simulation is corrected using the efficiency measured using \piz decays in data. 
The systematic uncertainty is estimated by varying independently for each \ptg bin the converted photon efficiency within the 
measurement uncertainty and computing the corrected ratio of efficiency $\varepsilon^{\g}_{\chicone}/\varepsilon^{\g}_{\chictwo}$ for each \ptjpsi bin. 
The systematic uncertainty is defined as the maximum variation observed.
The correction and the systematic uncertainty due to the \jpsi selection and reconstruction efficiency are found to be negligible.

The efficiency can be affected by the choice of the simulated \chic  \pt  spectrum (\ptchic): since the photon transverse momentum is correlated with the \jpsi transverse momentum,
the efficiency for each \ptjpsi bin can vary depending on the \ptjpsi spectrum inside this bin. 
In order to assess the uncertainty due to the \pt spectrum shape, the simulated \chictwo (\chicone) spectrum is changed to be 
identical to the simulated \chicone (\chictwo) \pt spectrum. 
The generated \chictwo and \chiczero decays have the same \pt dependence. For the ratio of \chiczero to \chictwo cross-sections 
the systematic uncertainty is assessed using the \pt spectrum of the \chicone mesons instead (alternatively for \chictwo or \chiczero mesons): 
the efficiency ratio varies by $\pm13\%$.

All of the systematic uncertainties are uncorrelated among bins, except those related to the \pt spectrum shape.
Table~\ref{tbl:systematics} summarises the systematic uncertainties on the ratio of yields for each \ptjpsi bin.

The ratio of cross-sections is also affected by the uncertainties on the branching fraction of $\chic\to\jpsi\g$ leading to
an additional systematic uncertainty of $6.0\%$ ($8.0\%$) on the cross section ratio $\sigma(\chictwo)/\sigma(\chicone)$ ($\sigma(\chiczero)/\sigma(\chictwo)$).  For each \ptjpsi bin the total systematic uncertainty is defined as the quadratic sum of all the systematic uncertainties detailed here. 
\begin{table}
\caption{\label{tbl:systematics}\small
Systematic uncertainties on the ratio of \chictwo and \chicone yields  for each \ptjpsi bin (in percent). 
The total systematic uncertainty is defined as the quadratic sum of all the systematic uncertainties.}
\begin{center}\begin{tabular}{l c c c c c c c c c c c}
\hline
\ptjpsi bin (\gevcnosp)        & 3-4 & 4-5 & 5-6 & 6-7 & 7-8 & 8-9 & 9-11 & 11-13 & 13-16 & 16-20 & 4-20\\
\hline
Fit bias                   & 1.8 & 2.0 & 3.0 & 3.0 & 3.0 & 3.0 & 3.0  & 3.0   & 3.0  & 3.0  & 3.0\\
Fit                        & 2.6 & 4.0 & 2.2 & 2.0 & 2.0 & 2.2 & 2.0  & 2.8   & 5.5  & 4.0  & 2.0\\
Comb bkg                   & 1.0 & 1.0 & 1.0 & 1.0 & 1.0 & 1.0 & 1.0  & 1.0   & 1.0  & 1.0  & 1.0 \\
Peaking bkg                & 0.3 & 0.3 & 0.3 & 0.3 & 0.3 & 0.3 & 0.4  & 0.4  & 0.4  & 0.4 & 0.4 \\
Photon efficiency          & 4.0 & 4.0 & 4.0 & 4.0 & 4.0 & 4.0 & 4.0  & 4.0  & 4.0  & 4.0 & 2.0\\
\ptchic spectrum           & 2.6& 2.4 & 2.2 & 2.1 & 2.0 & 1.8 & 1.6   & 1.3  & 1.0  & 0.7 & 6.4\\
\hline
Total                      & 5.8 & 6.5 & 6.0 & 5.9 & 5.8 & 5.8 & 5.7 & 6.0  & 7.6 & 6.5 & 8.2\\
\hline
\end{tabular}
\end{center}
\end{table}

\section{$\boldsymbol{\chic}$ polarization}\label{sec:polarization}
The prompt \chic polarization is unknown. The simulated \chic mesons are unpolarized and all the efficiencies given in the previous sections are therefore
determined under the assumption that the \chicone and the \chictwo mesons are produced unpolarized. The photon and \jpsi momentum distributions depend on the  
polarization of the \chic state and the same is true for the ratio of efficiencies. 
The correction factors for the ratio of efficiencies under other polarization scenarios are derived here.

The angular distribution of the $\chic\to\jpsi\gamma$ decay is described by the angles $\theta_{\jpsi}$, $\theta_{\chic}$ and $\phi$ where: $\theta_{\jpsi}$ is the angle between the directions
of the positive muon in the \jpsi rest frame and the \jpsi in the \chic rest frame; $\theta_{\chic}$ is the angle between the directions of the \jpsi in 
the \chic rest frame and the \chic in the laboratory frame; $\phi$ is the angle between the \jpsi decay plane in the \chic rest frame and 
the plane formed by the \chic direction in the laboratory frame and the direction of the \jpsi in the \chic rest frame.
The angular distributions of the \chic states depend on $m_{\chi_{cJ}}$, which is the azimuthal angular momentum quantum number of the $\chi_{cJ}$
state.
The general expressions for the angular distributions are independent of the choice of polarization axis (here chosen as
the direction of the \chic in the laboratory frame) and are detailed in Ref.~\cite{HERABchic}.
For each simulated event in the unpolarized sample, a weight is calculated from the values of $\theta_{\jpsi}$, $\theta_{\chic}$ and $\phi$ in the various 
polarization hypotheses and the ratio of efficiencies is deduced for each ($m_{\chi_{c1}}$,$m_{\chi_{c2}}$) polarization combination.
Table~\ref{tbl:polcor} gives the correction factors to apply to the final $\sigma(\chictwo)/\sigma(\chicone)$ results   for each ($m_{\chi_{c1}}$,$m_{\chi_{c2}}$) polarization combination. 

These corrections are different from those found in the analysis using calorimetric photons [12]. This is due to
the fact that the acceptance efficiency of converted photons highly depends on the polar angle of the photon: for
large angles there is a higher probability that one of the electrons escapes the detector before the calorimeter.
The systematic uncertainties estimated in the case where both \chicone and \chictwo mesons are produced unpolarized
also apply to the other polarization scenarios.
\begin{table}
\caption{\label{tbl:polcor} \small 
Correction factors to be applied to the final $\sigma(\chictwo)/\sigma(\chicone)$ results for each \ptjpsi bin for different combinations of 
\chicone and \chictwo polarization states $|J,m_{\chi_{cJ}}>$ with 
$|m_{\chi_{cJ}}|=0,...,J$ (``unpol'' means the \chic is unpolarized). 
The polarization axis is defined as the direction of the \chic in the laboratory frame.}
\begin{center}
\begin{tabular}{l c c c c c c c c c c}
\hline
  &\multicolumn{10}{c}{\ptjpsi [\gevc]}\\
($|m_{\chi_{c1}}|$,$|m_{\chi_{c2}}|$) & 3-4 & 4-5 & 5-6 & 6-7 & 7-8 & 8-9 & 9-11 & 11-13 & 13-16 & 16-20\\
\hline
(unpol,0) & 1.07& 1.04& 1.00& 0.96& 0.93& 0.94& 0.91& 0.87& 0.89& 0.86 \\
(unpol,1) & 0.99& 0.99& 0.98& 0.98& 0.98& 0.98& 0.97& 0.96& 0.95& 0.98 \\
(unpol,2) & 0.97& 0.98& 1.02& 1.05& 1.08& 1.07& 1.13& 1.16& 1.16& 1.16 \\
(0,unpol) & 1.03& 1.01& 0.98& 0.97& 0.94& 0.92& 0.94& 0.91& 0.89& 0.90 \\
(0,0) & 1.10& 1.05& 0.98& 0.93& 0.88& 0.86& 0.85& 0.79& 0.79& 0.77 \\
(0,1) & 1.02& 1.00& 0.96& 0.95& 0.92& 0.90& 0.90& 0.88& 0.84& 0.88 \\
(0,2) & 1.00& 0.99& 1.00& 1.01& 1.02& 0.98& 1.06& 1.05& 1.03& 1.05 \\
(1,unpol) & 1.00& 1.01& 1.02& 1.02& 1.03& 1.03& 1.04& 1.06& 1.05& 1.07\\ 
(1,0) & 1.07& 1.05& 1.02& 0.98& 0.96& 0.97& 0.94& 0.92& 0.93& 0.92 \\
(1,1) & 0.99& 1.00& 1.00& 1.00& 1.01& 1.01& 1.00& 1.02& 1.00& 1.05 \\
(1,2) & 0.97& 0.98& 1.04& 1.06& 1.11& 1.11& 1.17& 1.22& 1.22& 1.25 \\
\hline
\end{tabular}
\end{center}
\end{table}

\section{Results}
\begin{figure}[tb]
  \begin{center}
    \begin{tabular}{cc}
      \hspace{-.5cm}\includegraphics[width=0.52\linewidth]{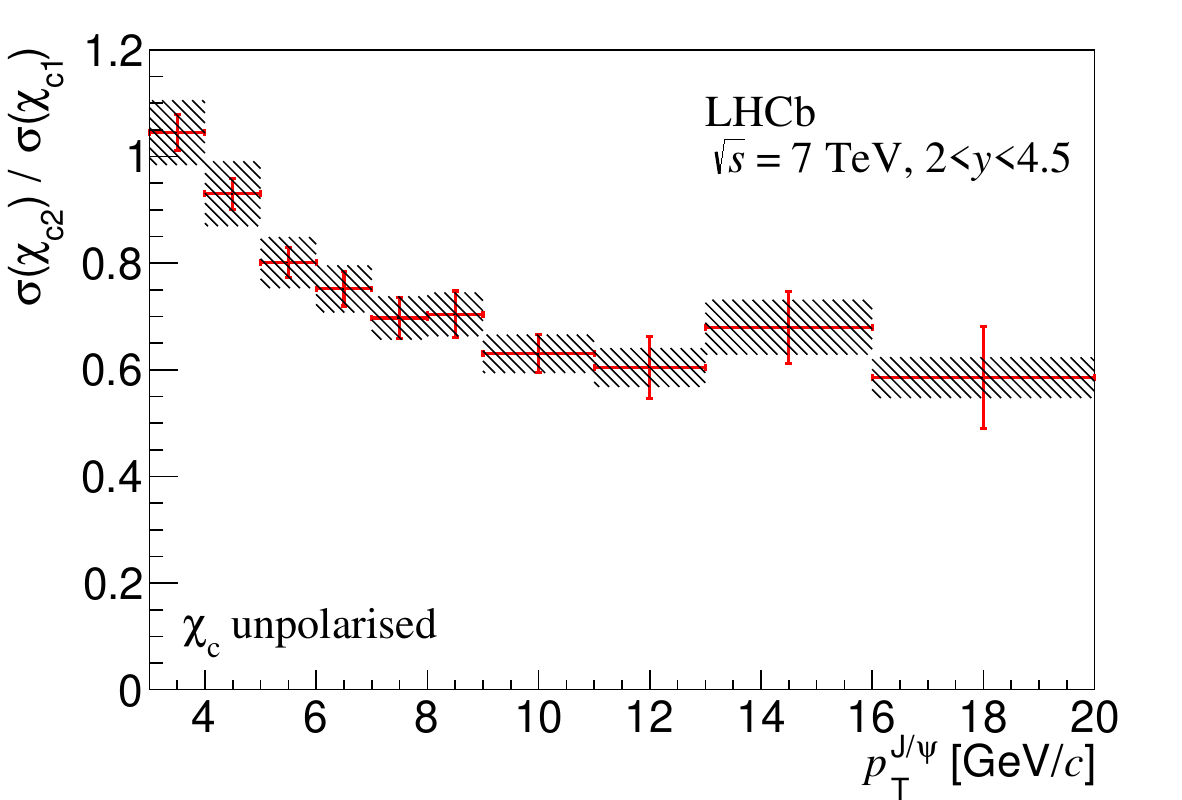} & 
      \hspace{-.5cm}\vspace{-.5cm}\includegraphics[width=0.52\linewidth]{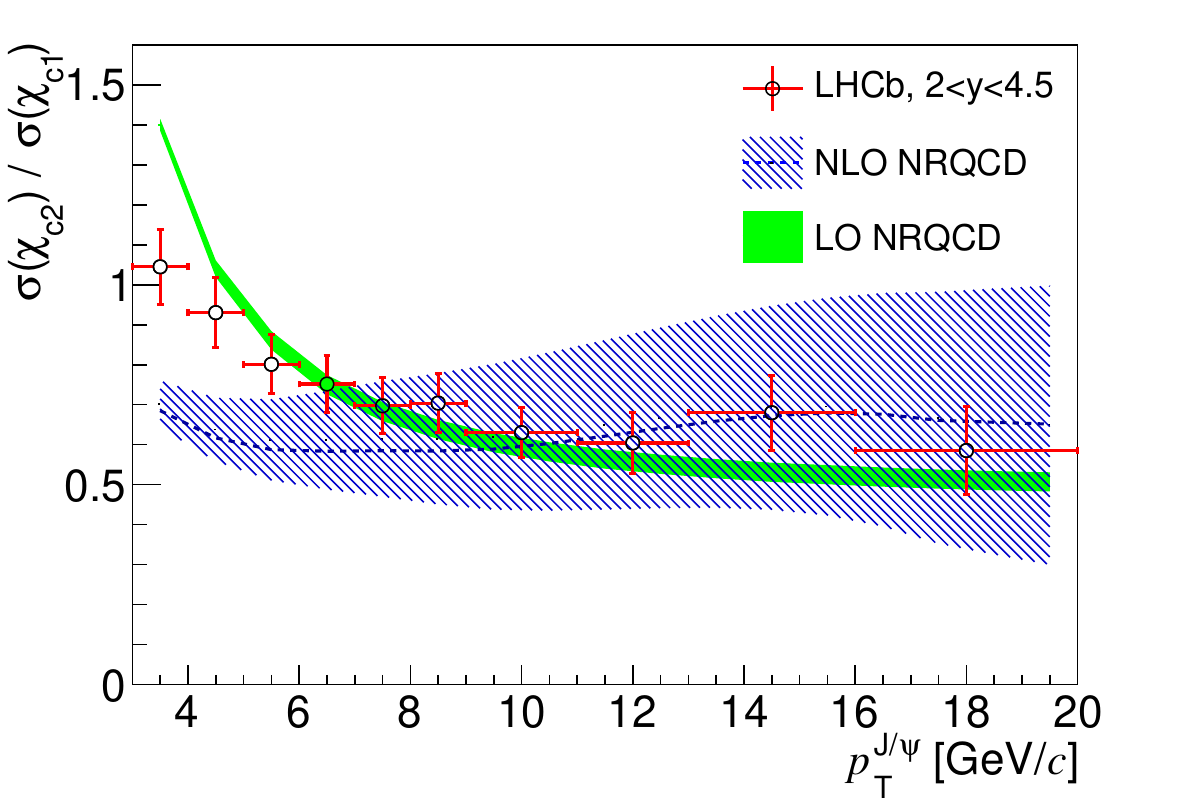}
    \end{tabular}
  \end{center}
  \caption{
    \small (left) Ratio of \chictwo to \chicone cross-sections  at $\sqrt{s}=7$~\tev for $2.0<y<4.5$. 
The statistical uncertainty is shown with a red error bar and the systematic uncertainty with a hashed rectangle.
(right) Comparison of the LHCb results (with total uncertainty) with the NLO NRQCD calculation from Ref.~\cite{NRQCD} (blue shading) and the LO 
NRQCD calculation of Ref.~\cite{Likhoded_chic} (solid green). The LHCb results are obtained assuming the \chic mesons are produced unpolarized.
  \label{fig:sigmaratio}}
\end{figure}

For each \ptjpsi bin the ratio of \chictwo to \chicone yields, obtained from a least squares fit described in Sec~\ref{sec:fit},
 is corrected for the peaking background (see Sec.~\ref{sec:bkg}), by the efficiency ratio (see Sec.~\ref{sec:effic})
and by the ratio of branching fractions of $\chic\to\jpsi\gamma$ (see Sec.~\ref{sec:signal}).
Figure~\ref{fig:sigmaratio} (left) shows the ratio of the \chictwo to \chicone production cross-sections as a function of \ptjpsi
 under the assumption that the \chic mesons are produced unpolarized. 
The overall systematic uncertainty ($6.0\%$) due to the branching fraction of $\chic\to\jpsi\gamma$ is not shown here.
Table~\ref{tbl:results} gives the ratio of cross-sections with their statistical and systematic uncertainties for each \ptjpsi bin
including that originating from the unknown polarization of the \chic states.
Figure~\ref{fig:sigmaratio} (right) shows  a comparison of this measurement with the next to leading order (NLO) NRQCD calculation of Ref.~\cite{NRQCD} and with the 
LO NRQCD calculation of Ref.~\cite{Likhoded_chic}.

A \chiczero signal is observed for $4<\ptjpsi<20\gevc$ with a statistical significance, determined from the ratio of the signal yield and its 
uncertainty, of $4.3~\sigma$  and the extracted yield is $N(\chiczero)=705\pm163$. 
The ratio of \chiczero and \chictwo yields obtained from the fit is corrected by the efficiency ratio (see Sec.~\ref{sec:effic}) 
and the ratio of branching fractions in order to obtain
the ratio of cross-sections (under the hypothesis of unpolarized states) and integrated over \ptjpsi
\begin{equation*}
\sigma(\chiczero)/\sigma(\chictwo)=1.19\pm0.27\,\stat \pm0.29\,\syst\pm0.16\,(\,\pt\,\mathrm{model})\pm0.09\,(\,\BR),
\end{equation*}
where the first uncertainty is statistical, the second is the systematic uncertainty dominated by the photon 
efficiency, the \chicone tail parameters and background modelling, the third from the choice of \pt spectrum 
and the fourth from the branching fraction uncertainty. 
For comparison, the ratio of \chictwo to \chicone production cross-sections for the same \ptjpsi range is
\begin{equation*}
\sigma(\chictwo)/\sigma(\chicone)=0.787\pm0.014\,\stat \pm0.034\,\syst \pm0.051\,(\,\pt\, \mathrm{ model})\pm0.047\,(\,\BR).
\end{equation*}
\begin{figure}[tb]
  \begin{center}
\begin{tabular}{cc}
\hspace{-.5cm}\includegraphics[width=0.52\linewidth]{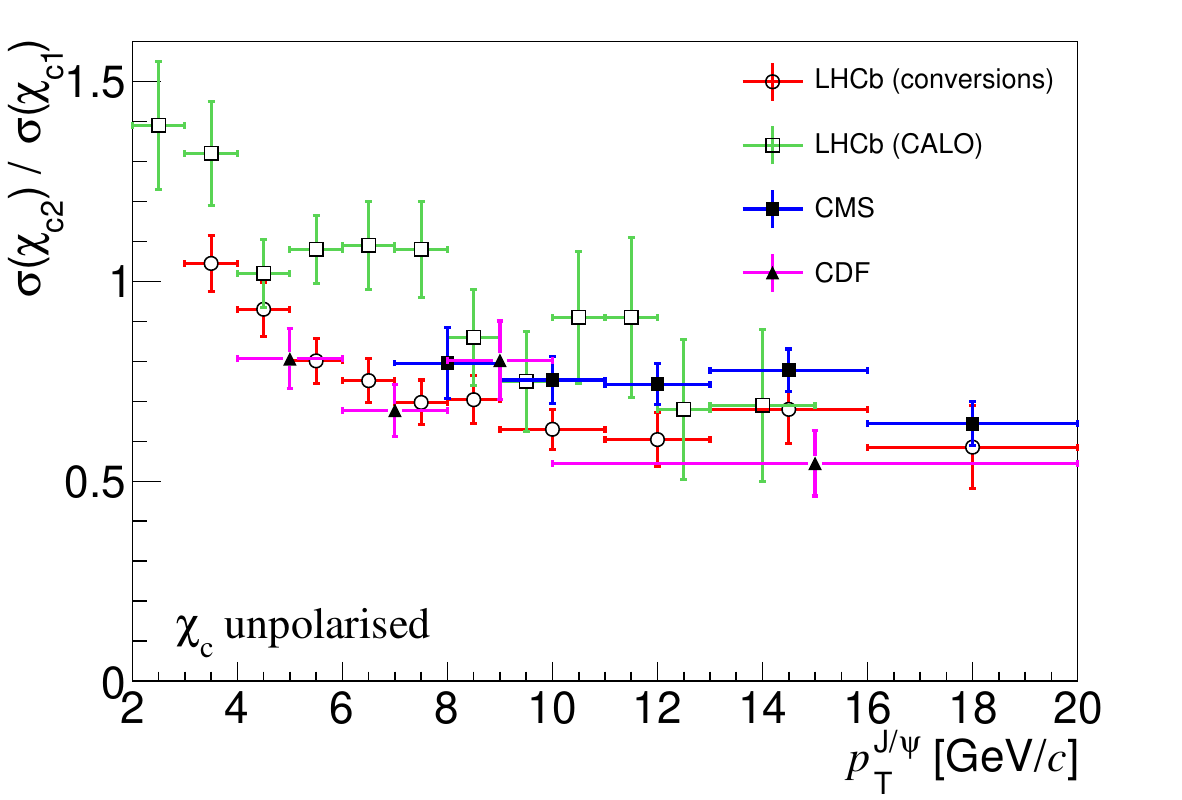} &
\hspace{-.5cm}\includegraphics[width=0.52\linewidth]{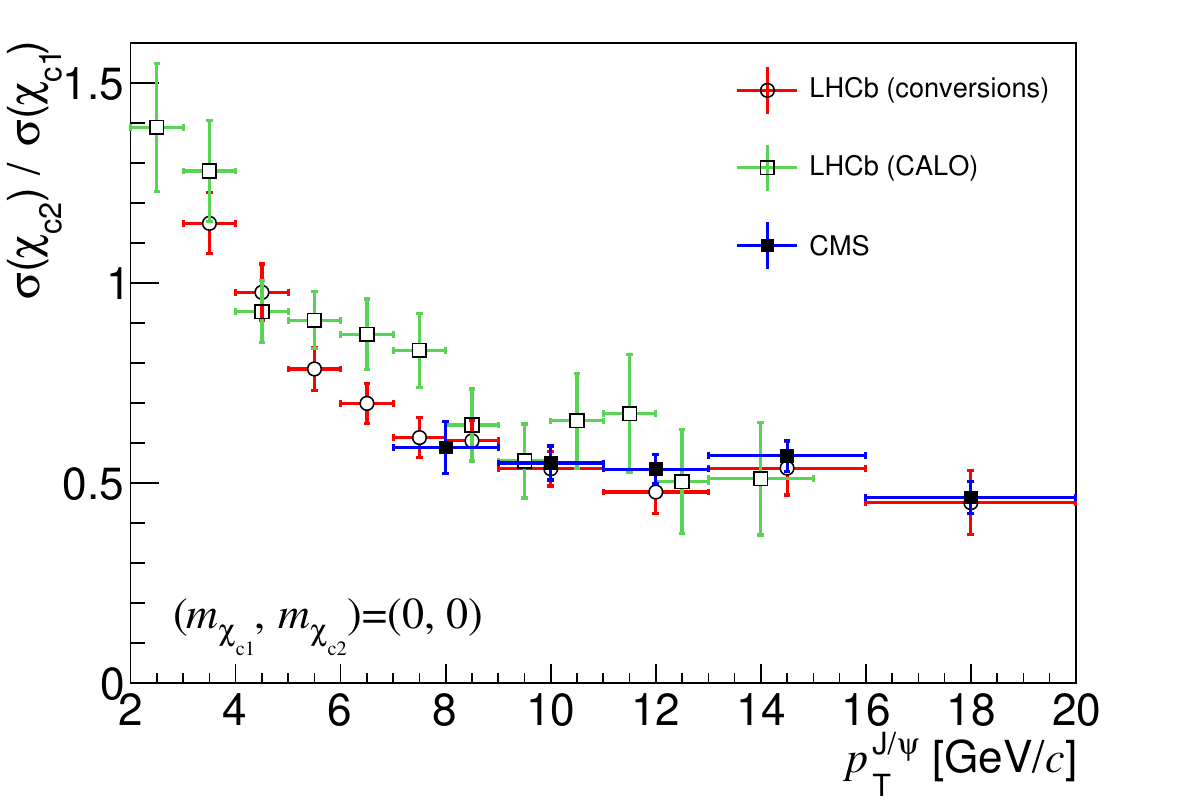}\\
\end{tabular}
  \end{center}
  \caption{
    \small Comparison of the ratio of \chictwo to \chicone cross-sections obtained by LHCb using calorimetric photons~\cite{LHCb-PAPER-2011-019} (green open squares), 
CMS result~\cite{CMSchic} (blue filled squares), CDF result (purple filled triangles)~\cite{CDFchic} and the result presented here (red open circles)
under the assumption (left) of unpolarized states and (right) under the assumption  $(m_{\chicone},m_{\chictwo})=(0,0)$ in the helicity frame.
The uncertainty due to the limited knowledge of the branching fractions of $\chic\to\jpsi\gamma$, which is common to all the measurements, is not included here.}
  \label{fig:ratiocomp}
\end{figure}
\begin{table}
\caption{\small Measurements of the ratio of \chictwo to \chicone production cross-sections for the given \ptjpsi range assuming unpolarized \chic production.
The first uncertainty is statistical, the second is systematic, the third is from the branching fractions used
and the last gives the maximum correction due to the unknown polarization.\label{tbl:results} }
\begin{center}
\setlength{\extrarowheight}{2mm}
\begin{tabular}{cccc}
\hline
\ptjpsi[\gevcnosp] & $\sigma(\chictwo)/\sigma(\chicone)$ \\
\hline
$3-4$ & $1.037 \pm 0.033\stat  \pm 0.060\syst \pm 0.062\,(\BR)\, ^{+0.10}_{-0.03} ({\rm pol})$ \\
$4-5$ & $0.923 \pm 0.029\stat  \pm 0.060\syst \pm 0.055\,(\BR)\, ^{+0.05}_{-0.02} ({\rm pol})$ \\
$5-6$ & $0.795 \pm 0.028\stat  \pm 0.048\syst \pm 0.048\,(\BR)\, ^{+0.03}_{-0.03} ({\rm pol})$ \\
$6-7$ & $0.746 \pm 0.032\stat  \pm 0.044\syst \pm 0.045\,(\BR)\, ^{+0.05}_{-0.05} ({\rm pol})$ \\
$7-8$ & $0.692 \pm 0.039\stat  \pm 0.040\syst \pm 0.042\,(\BR)\, ^{+0.08}_{-0.08} ({\rm pol})$ \\ %
$8-9$ & $0.699 \pm 0.044\stat  \pm 0.041\syst \pm 0.042\,(\BR)\, ^{+0.08}_{-0.10} ({\rm pol})$ \\
$\,\,9-11$ & $0.625 \pm 0.035\stat  \pm 0.036\syst \pm 0.038\,(\BR)\, ^{+0.11}_{-0.09} ({\rm pol})$ \\
$11-13$ & $0.600 \pm 0.057\stat  \pm 0.036\syst \pm 0.036\,(\BR)\, ^{+0.13}_{-0.13}  ({\rm pol})$ \\ %
$13-16$ & $0.675 \pm 0.067\stat  \pm 0.051\syst \pm 0.040\,(\BR)\, ^{+0.15}_{-0.15} ({\rm pol})$ \\ %
$16-20$ & $0.581 \pm 0.096\stat  \pm 0.038\syst \pm 0.035\,(\BR)\, ^{+0.15}_{-0.15} ({\rm pol})$ \\ %
\hline
\end{tabular}
\end{center}
\end{table}
%
\section{Conclusion}
The ratio of prompt production cross-sections of \chictwo and \chicone is measured in a rapidity range $2.0<y<4.5$ as a function of \ptjpsi from 3 to 20 \gevc at 
$\sqrt{s}=7$ \tev using the decays $\chic\to\jpsi\g$ where the photon converts in the detector material. 

This ratio was also measured by LHCb using calorimetric photons~\cite{LHCb-PAPER-2011-019}, by the 
CMS experiment~\cite{CMSchic} in the rapidity range $|y|<1$ using converted photons at $\sqrt{s}=7$~\tev 
and by CDF~\cite{CDFchic} using converted photons at $\sqrt{s}=1.96$~\tev in the range $|\eta(\jpsi)|<1$ and $\pt(\g)>1.0$\gevc. 
These measurements are compared in Fig.~\ref{fig:ratiocomp}. The ratios are expected to be similar for $pp$ and ${\ensuremath{\Pp}}\antiproton$
 collisions
since \chic mesons are  produced predominantly via gluon-gluon interactions and depend only  weakly on the centre-of-mass energy and $y$ coverage~\cite{NRQCD,Likhoded_chib}.
The results from this analysis are compatible with the CMS and CDF results. 
The statistical and systematic 
uncertainties can be safely assumed to be uncorrelated between the analysis presented here and the LHCb analysis using calorimetric photons, since 
the data samples are different, the photon reconstruction is based on different subdetectors (calorimeter or tracker) and the background modelling 
is performed in a  different way.
The measurements are in agreement but the results of the analysis using converted photons are systematically lower. 
As underlined in Sec.~\ref{sec:polarization} analysis-dependent corrections have to be applied to these ratios depending on the 
polarization hypothesis (see Table~\ref{tbl:polcor}). 
When correcting the results assuming the \chic states are polarized with $(m_{\chicone},m_{\chictwo})=(0,0)$, all the results are in better agreement
as shown in Fig.~\ref{fig:ratiocomp} (right).

The \chiczero meson prompt production is also studied and its production cross section ratio relative to  the \chictwo meson 
is measured in the range $4\gevc<\ptjpsi<20\gevc$. 
This is the first evidence for \chiczero meson production at a hadron collider. 
Our result is in agreement with the NLO NRQCD prediction of $\sigma(\chiczero)/\sigma(\chictwo)=0.62\pm0.10$ ($4<\ptjpsi<20\gevc$)~\cite{NRQCD}
and with the LO NRCQD prediction of $\sigma(\chiczero)/\sigma(\chictwo)=0.53\pm0.02$ ($4<\ptjpsi<20\gevc$)~\cite{Likhoded_chic}.

\section*{Acknowledgements}
\noindent 
We thank Y.Q. Ma for providing the NLO NRQCD predictions. We also thank 
A. Luchinsky and A. Likhoded for providing the 
LO NRQCD predictions and for interesting discussions.
We express our gratitude to our colleagues in the CERN
accelerator departments for the excellent performance of the LHC. We
thank the technical and administrative staff at the LHCb
institutes. We acknowledge support from CERN and from the national
agencies: CAPES, CNPq, FAPERJ and FINEP (Brazil); NSFC (China);
CNRS/IN2P3 and Region Auvergne (France); BMBF, DFG, HGF and MPG
(Germany); SFI (Ireland); INFN (Italy); FOM and NWO (The Netherlands);
SCSR (Poland); MEN/IFA (Romania); MinES, Rosatom, RFBR and NRC
``Kurchatov Institute'' (Russia); MinECo, XuntaGal and GENCAT (Spain);
SNSF and SER (Switzerland); NAS Ukraine (Ukraine); STFC (United
Kingdom); NSF (USA). We also acknowledge the support received from the
ERC under FP7. The Tier1 computing centres are supported by IN2P3
(France), KIT and BMBF (Germany), INFN (Italy), NWO and SURF (The
Netherlands), PIC (Spain), GridPP (United Kingdom). We are thankful
for the computing resources put at our disposal by Yandex LLC
(Russia), as well as to the communities behind the multiple open
source software packages that we depend on.

\addcontentsline{toc}{section}{References}
\setboolean{inbibliography}{true}
\bibliographystyle{LHCb}
\bibliography{main,LHCb-PAPER,LHCb-DP}

\end{document}